\documentclass[twocolumn,showpacs,preprintnumbers,amsmath,amssymb]{revtex4}

\usepackage{graphicx}
\usepackage{dcolumn}
\usepackage{bm}

\newcommand{\tr } {\mathrm{tr} }
 
\newtheorem{lemma}{Lemma}

\newcommand{\proof } {\noindent{\bf{Proof.}} }

\begin{document}

\preprint{APS/123-QED}

\title{Estimation of $n$ non-identical unitary channels}

\author{Caleb O'Loan}
 \email{cjo2@st-andrews.ac.uk}
\affiliation{%
School of Mathematics and Statistics, University of St Andrews, KY16 9SS
}%
 \homepage{http://www.st-andrews.ac.uk/~cjo2}

\date{\today}

\begin{abstract}
We investigate the simultaneous estimation of $n$ not necessarily identical unitary channels using multi-partite entanglement. We examine whether it is possible for the rate at which the mean square error decreases to be greater than that using the channels individually.
For a reasonably general situation, in which there is no functional dependence between the channels, we show that this is not possible.
We look at a case in which the channels are not necessarily identical but depend on a common variable. 
In this case, the mean square error decreases more rapidly using multi-partite entanglement.

\end{abstract}

\maketitle

{ \em  Introduction.}
Estimation of quantum states and channels is of fundamental importance to quantum information theory. 
Estimation of unitary channels, when $n$ copies are available, has received a lot of attention; for an accessible overview see \cite{zhen06}.  Many schemes have been devised for which the error decreases at a much faster rate, compared to a straightforward approach of using each channel separately \cite{haysahi06, kahn07}. This is analogous to the problem of transmitting a reference frame \cite{baganetal04a,baganetal04b,chiribella04,rudolph03}.
This increase in the rate of estimation is possible both with \cite{haysahi06, kahn07,baganetal04a,baganetal04b,chiribella04} and without \cite{burghbart05,rudolph03} the use of entanglement.
 
 As far as we are aware, no work has been done on the situation in which there are $n$ non-identical channels. In this paper we investigate whether an increase in the rate of estimation is possible in this case. We look at the following two cases: (i) there are $n$ unitary channels belonging to $SU(d)$ specified by the parameters $\theta^1, \dots, \theta^n$, (ii) there are $n$ not necessarily identical one-parameter unitary channels which depend on a common variable.
In the former scheme an increase in the rate of estimation is not possible; in the latter scheme it is.

To quantify the performances of estimation schemes we use the asymptotic limit of the mean square error $E[(\theta^i - \hat \theta^i)(\theta^j - \hat \theta^j)]$. Using the maximum likelihood estimator, as the number of measurements $N \rightarrow \infty$, the mean square error is approximately $(1/N) F_M^{-1}(\theta)$ \cite{hayashi06b}, where $F_M(\theta)$ is the Fisher information matrix obtained from a single measurement $M$.
The Fisher information for the parameter $\theta = (\theta^1, \dots, \theta^p)$ is the $p \times p$ matrix with entries
\begin{eqnarray*}
 F_{M}(\theta)_{jk} &\equiv&  \int  p(\xi;\theta) \left( \frac{\partial \ln p(\xi;\theta)}{\partial \theta^j} \right)\left( \frac{\partial \ln p(\xi;\theta)}{\partial \theta^k} \right) d\xi . 
\label{eq:FI00}
\end{eqnarray*}
 The Cram\'er--Rao inequality states that the mean square error of an unbiased estimator $t(x)$ is less than or equal to the inverse of the Fisher information,
\begin{equation}
\mathrm{m.s.e.}_\theta[t(x)] \geq F_M(\theta)^{-1}.
\label{eq.crineq}
\end{equation}
It has been shown  \cite{braunsteincaves94} that the SLD quantum information $H(\theta)$ is an upper bound on the Fisher information, i.e. $F_M(\theta) \leq H(\theta)$.
The SLD quantum information for the parameter $\theta = (\theta^1, \dots, \theta^p)$ is defined as the matrix with entries
\begin{equation*}
H_{jk} = \Re \tr \{ \lambda^j \rho \lambda^k \},
\end{equation*}
where $\lambda^j$ is any self-adjoint solution to the matrix equation
\begin{equation}
\frac{d \rho}{d \theta^j} = \frac{1}{2}(\rho \lambda^j + \lambda^j \rho).
\label{eq.qmscore}
\end{equation}
Given a state $\rho_\theta$ depending on some unknown parameter $\theta$, and a POVM $\{ M_m \}$ we get a measurement outcome $x$.
 The quantum Cram\'er--Rao inequality states that the mean square error of an unbiased estimator $t(x)$ of $\theta$ is less than or equal to the inverse  of the SLD quantum information, i.e.
\begin{equation}
\mathrm{m.s.e.}_{\theta}[t(x)] \geq H^{-1}(\theta).
\label{eq.qcr}
\end{equation}
When $p \geq 2$, there exist families of states $\rho_\theta$ and $\sigma_\theta$  for which neither $H(\rho_\theta) \geq H(\sigma_\theta)$ or $H(\rho_\theta) \leq H(\sigma_\theta)$ is true.
To deal with this, we compare the traces $\tr \{H(\theta)\}$ of the SLD quantum information.

Previous work has looked at how the error scales with the number of times $U$ is used. We look at how the error scales with the number of input states used, as this is more convenient for us.

\section{`Independent' Channels}
First we look at estimating $n$ unitary channels from $SU(d)$. The $j$th channel is specified by the parameter $\theta^{j}=(\theta^j_1, \dots, \theta^j_{d^2-1})$. 
These channels are supposed `independent' in that there is no functional relationship between $\theta^1, \theta^2, \dots, \theta^{n-1}$ and $\theta^{n}$.
Estimation of the channels is a parametric problem.

We look at the mapping
\begin{equation}
\rho_0 \mapsto \otimes_{j=1}^n (U_{\theta^j} \otimes\mathbb{I}_R) \rho_0 (U_{\theta^j}^\dagger \otimes\mathbb{I}_R). 
\label{eq.indch}
\end{equation}
We show in Appendix A that using a tensor product of maximally entangled input states $\rho_0 = \otimes_{j=1}^n \rho_{mes}^j$ is sufficient to maximize the trace of the SLD quantum information of the output states of (\ref{eq.indch}).
This is not a necessary condition, as the $2n$-partite entangled state $1/\sqrt{d} \sum_i | e_i \rangle \otimes \cdots \otimes | e_i \rangle$ also attains the maximum SLD quantum information. However, since these states are significantly harder to produce, we are better off using maximally entangled states.
The SLD quantum information using a tensor product of maximally entangled states is attainable \cite{ballestera}, so asymptotically the mean square error is $(1/N) H^{-1}(\theta)$.

For the mapping (\ref{eq.indch}), an optimal estimation procedure for $n$ `independent' unitary channels, in terms of $\tr \{H\}$,  is to estimate each one individually using a maximally entangled input state. The optimality of this procedure, in terms of $\tr\{H(\theta)\}$ has been shown by Ballester \cite{ballestera}.

\section{`Dependent' Channels}

We look at the case where we have $n$ not necessarily identical channels which depend on a common parameter $\theta$. These channels are of the form
\begin{equation}
U^{1}_\theta = \left( \begin{array}{cc}
1 & 0 \\
0 & e^{i f_1(\theta)} \
\end{array} \right),
\dots, \quad 
U^{n}_\theta = \left( \begin{array}{cc}
1 & 0 \\
0 & e^{i f_n(\theta)} \
\end{array} \right),
\label{eq.u12un}
\end{equation}
where $0 \leq \theta \leq t$ and $\mathcal{H} = \mathbb{C}^2$. We can use each channel $N$ times.
 We impose the following conditions on the functions $f_j$: (a) $f_j(\theta): \mathbb{R} \mapsto  \mathbb{R}$,  (b)   $df_j(\theta)/d\theta \geq 0$, and (c) $0 \leq \sum_j f_j(\theta) \leq  \pi$, for all $j$ and $\theta$.
We look at the mapping
\begin{equation}
\rho_0 \mapsto \otimes_{j=1}^n( U_\theta^j \otimes \mathbb{I}_R) \rho_0 ( U_\theta^{j \dagger} \otimes \mathbb{I}_R),
\label{eq.sch2}
\end{equation}
where $\rho_0 \in S(\otimes^n (\mathcal{H} \otimes \mathcal{H}_R))$; we denote by $S(\mathcal{H})$ the set of states on $\mathcal{H}$. We compare the SLD quantum information between (i) using a tensor product of maximally entangled states, (ii) using a $2n$-partite entangled state.
In this case using $2n$-partite entanglement gives considerably larger SLD quantum information. 

Using the input state $\rho_0 = \rho_{mes}^1 \otimes \cdots \otimes \rho_{mes}^n, $ where $\rho_{mes}^j  = | \psi_u^j \rangle \langle \psi_u^j |$, $| \psi_u^j \rangle =   1/\sqrt{2}(| 00 \rangle + | 11 \rangle )$, gives an SLD quantum information of $\sum_{j=1}^n (df_j/d\theta)^2$, which is attainable using a tensor product of the POVM $\{ M_0^j = | \psi_u^j \rangle \langle \psi_u^j |, M_1^j = \mathbb{I} - | \psi_u^j \rangle \langle \psi_u^j | \}$. 

If we use the input state $\rho_0 = | \psi_0 \rangle \langle \psi_0 |$, where $| \psi_0 \rangle = 1/\sqrt{2} (| 0 0 .. 0 \rangle + | 1 1 .. 1 \rangle)$, we get an SLD quantum information of $(\sum_{j=1}^n df_j/d\theta)^2$. As  $df_j/d\theta \geq 0$ for all $j$,  this is considerably larger than the SLD quantum information using a tensor product of maximally entangled states.

The phase $\phi$ of the output state is in one-to-one correspondence with $\theta$,  because of conditions (b) and (c); in fact $\phi = \sum_{j=1}^n f_j(\theta)$.
Using the POVM $\{ M_0 = | \psi_0 \rangle \langle \psi_0 |, M_1 =\mathbb{I} - | \psi_0 \rangle \langle \psi_0 | \}$ we get 
$p(0; \phi) = \cos^2(\phi/2)$.
We perform (\ref{eq.sch2}) $N$ times and each time use this POVM.  From the measurement outcomes we get an estimate $\hat \phi$ of $\phi$ by  $\cos^2(\hat \phi/2) = n_0/N$, where $n_0$ is the number of times we get the outcome $x=0$. Because of condition (c) on the functions $f_j(\theta)$, $\cos^2(\hat \phi/2)$ is in one-to-one correspondence with $\hat \phi$ and hence $\hat \theta$.
 This POVM gives a Fisher information equal to the SLD quantum information. Hence as $N  \rightarrow \infty$ we get a mean square error of $1/(N (\sum_{j=1}^n df_j/d\theta)^2)$ compared to $1/(N \sum_{j=1}^n (df_j/d\theta)^2)$ using a tensor product of maximally entangled input states.

It is known that when we have $n$ identical simple unitary channels, we can obtain an increase in the rate of estimation without using entanglement \cite{zhen06}. A simple way is to use each of the $n$ channels in sequence on a single input state. Consider the unitary channel $U_\theta = \mathrm{Diag}(1, e^{i\theta})$
, i.e.
\begin{equation}
\rho_0 \mapsto U_\theta^n \rho_0 U_\theta^{n \dagger}, \quad U_\theta^n = U_\theta U_\theta \cdots U_\theta. 
\label{eq.seq1}
\end{equation}
We repeat (\ref{eq.seq1}) $N$ times using the input state $\rho_x = | \psi_x \rangle \langle \psi_x |$ where $| \psi_x \rangle = 1/\sqrt{2}(| 0 \rangle + | 1 \rangle )$, and the POVM $\{ |\psi_x \rangle \langle \psi_x |, \mathbb{I}  - |\psi_x \rangle \langle \psi_x | \}$. We call this the {\em sequential scheme}. As $N \rightarrow \infty$, we obtain a mean square error that scales as $1/ (N n^2)$ . 
We can use this sequential scheme with $n$ non-identical channels, i.e.
\begin{equation}
\rho_0 \mapsto ( U_\theta^1 \cdots U_\theta^n) \rho_0 ( U_\theta^1 \cdots U_\theta^n)^\dagger,
\label{eq.sch3}
\end{equation}
where $\rho_0 \in S(\mathcal{H})$.
Using the input state $\rho_x = | \psi_x \rangle \langle \psi_x |$ where $| \psi_x \rangle = 1/\sqrt{2}(| 0 \rangle + | 1 \rangle )$, and the POVM $\{ |\psi_x \rangle \langle \psi_x |, \mathbb{I}  - |\psi_x \rangle \langle \psi_x | \}$ we get $p(0;\phi) = \cos^2(\phi/2)$.  Performing (\ref{eq.sch3}) $N$ times we get an estimate $\hat \phi$ and hence $\hat \theta$. We get the same Fisher information as (\ref{eq.sch2}) using a $2n$-partite entangled state, and hence the same mean square error.

Since multi-partite entanglement is difficult to create, we are better off using the sequential scheme (\ref{eq.sch3}).

  Conditions (b) and (c) for the functions $f_j(\theta)$ are very strict. We can still get a considerable increase in the rate of estimation without these conditions. If condition (c) does not hold, the phase $\phi$ of the output state is not in one-to-one correspondence with $p(0; \phi) = \cos^2(\phi/2)$. If we modify condition (c) to $0 \leq \sum_j f_j(\theta) \leq 2\pi$, we can still find $\phi$ but we shall need to perform an extra measurement $ \{ | \psi_y \rangle \langle \psi_y |, \mathbb{I} - | \psi_y \rangle \langle \psi_y | \}$, where $| \psi_y \rangle = 1/\sqrt{2}(| 0 \rangle +i | 1 \rangle )$, a small number of times to determine the sign of $\cos(\phi/2)$. Then we can estimate $\phi$ and hence $\theta$.
  If $0 \leq \sum_j f_j(\theta) \leq 2\pi$ does not hold, the phase $\phi$ of the output state is not in one-to-one correspondence with $\theta$.  We can get around this using a method similar to that of Zhengfeng { \it et al} \cite{zhen06}, based on Rudolph and Grover \cite{rudolph03}. 
 
 Zhengfeng {\it et al} \cite{zhen06} looked at the case where $f_j(\theta) = 2 \pi \theta$ for all $j$.
 Their scheme involves first using a single channel $n$ times to get an interval in which $\theta$ almost certainly lives. Then they use two or three channels in sequence to `amplify' $\theta$. This is repeated $n$ times until they get a narrower interval for $\theta$. This is continued until all channels are being used simultaneously. The mean square error scales as $\left( \log N' /N' \right)^2$, where $N'$ is the total number of times $U_\theta$ is used.
  
In our case, the situation is more complex, as the functions $f_j(\theta)$ are more general and not necessarily identical.   The finer details of how we go about this and how the mean square error would behave, depend on the functions $f_j(\theta)$. We give a very brief overview of a possible procedure.
We start by using a single unitary $U_1$ $n$ times to get an interval for $f_1(\theta)$ and hence $\theta$. Then we use $U_2 U_1$ to get an estimate of $f_1(\theta) + f_2(\theta)$ and hence a more accurate estimate of $\theta$.  We continue this process till we are using all the channels simultaneously and we have a very narrow interval for $\theta$. 
We expect that asymptotically, the mean square error is approximately  $1/(N (\sum_{j=1}^n df_j/d\theta)^2)$, where $N$ is the number of input states used. We leave a more in-depth analysis for further work.
  
  If neither (b) nor (c) are satisfied we propose the following scheme:
 (i) Use each of the channels individually $n$ times to get an estimate $\hat \theta$, 
  (ii) Divide the channels into two groups: $A = \{ U_\theta^j, df_j/d\theta \geq 0$ at $\hat \theta \}$,  $B = \{ U_\theta^k, df_k/d\theta < 0$ at $\hat \theta \}$,
  (iii) Use an iterative procedure for the two groups separately, but sharing information about $\theta$ to make the confidence intervals shorter.

\appendix
\section{Proof}
We are looking at unitary channels of the form \newline 
$U_\theta = \exp(i \sum_j \theta_j t_j)$ where $t_j = t_j^\dagger$, $\tr\{ t_j \} =0$ and $\tr\{ t_j t_k \} = \delta_{jk}$.
We denote by $H_{\theta^j}(\rho_{0})$, the SLD quantum information for the $j$th channel
\begin{equation*}
 \rho_{0} \mapsto (U_{\theta^j} \otimes \mathbb{I}) \rho_{0} (U_{\theta^j}^\dagger \otimes \mathbb{I}).
\label{eq.jthchan}
\end{equation*}
The SLD quantum information of (\ref{eq.indch}) using a tensor product of maximally entangled states is
 $H_\theta(\otimes_{j=1}^n \rho_{mes}) =  \mathrm{Diag}(H_{\theta^1}(\rho_{mes}), \dots,  H_{\theta^n}(\rho_{mes}))$.

\begin{lemma}
For all unitary channels of the form
\begin{equation}
\rho_0 \mapsto (U_\theta \otimes \mathbb{I})\rho_0(U_\theta^\dagger \otimes \mathbb{I}),
\label{eq.sepui}
\end{equation}
the trace of the SLD quantum information is maximized by a maximally entangled state,
i.e. 
\begin{equation}
\tr\{H_\theta(\rho_0)\} \leq \tr\{H_\theta(\rho_{mes})\}.
\label{lem.hsleqmes}
\end{equation}
Equality holds in (\ref{lem.hsleqmes}) if and only if $\rho_0$ is a maximally entangled state.
\end{lemma}
\proof
From the appendix of Ballester \cite{ballesterb} we know that for the channel (\ref{eq.sepui})
\begin{equation}
\tr \{ (H_\theta(\rho_{mes}))^{-1} H_\theta(\rho_0) \} \leq d^2-1.
\label{balres}
\end{equation}
It is simple to show that for unitary channels of the form $U_\theta = \exp(i \sum_j \theta_j t_j)$ we have  $H_\theta(\rho_{mes}) = (4/d) \mathbb{I}_{d^2-1}$. Substituting into (\ref{balres}) we get
\begin{equation}
\tr \{ H_\theta(\rho_0) \} \leq \frac{4 (d^2-1)}{d} = \tr \{ H_\theta(\rho_{mes}) \} .
\label{mybalres}
\end{equation}
Since equality holds in (\ref{balres}) if and only if $\rho_0$ is a maximally entangled state \cite{ballesterb}, equality holds in  (\ref{mybalres}) if and only if $\rho_0$ is a maximally entangled state.

\begin{lemma}
The trace of the SLD quantum information for (\ref{eq.indch}) is maximized by a tensor product of maximally entangled states , i.e. 
\begin{eqnarray}
\tr\{ H_{\theta}(\rho_0) \} \leq \tr\{H_{\theta}(\otimes_{j=1}^n\rho^j_{mes})\}, \label{eqn.htleqhi2} \\
 \rho_0 \in S(\otimes^n (\mathcal{H}\otimes \mathcal{H}_R)), \quad \rho^j_{mes} \in S( \mathcal{H}\otimes \mathcal{H}_R). \nonumber
\end{eqnarray}
\label{thm.htleqhi2}
\end{lemma}
For pure states a solution of (\ref{eq.qmscore}) is $\lambda^j = 2 d \rho_\theta/d \theta^j$. It is not difficult to show that for the set of states $U_\theta \rho_0 U_\theta^\dagger$ the SLD quantum information is the matrix with entries
\begin{eqnarray}
H_{\theta_{jk}}(\rho_0) &=& 4 \Re \tr \{ U_\theta^j \rho_0 U_\theta^{k \dagger} \} \nonumber \\
 &+& 4  \tr \{ U_\theta^j \rho_0 U_\theta^{ \dagger} \}  \tr \{ U_\theta^k \rho_0 U_\theta^{ \dagger} \} ,
\label{eq.sldjk}
\end{eqnarray}
where $U_\theta^j = \partial_{\theta^j} U_\theta$. In  (\ref{eq.indch})  $U_\theta = \otimes_{j=1}^n U_{\theta^j} \otimes \mathbb{I}_R$. 

Consider an arbitrary diagonal element of the SLD quantum information, i.e.\ $H_{\theta_{mm}}$, where $m(j,k) = (j-1)p + k$, corresponding to the parameter $\theta^j_k$. From (\ref{eq.sldjk}),
 \begin{equation*}
H_{\theta_{mm}}(\rho_0) = 4 \tr ( U_\theta^{j_k} \rho_0 U_\theta^{j_k \dagger}) 
+ 4\left(\tr (  U_\theta^{j_k}  \rho_0  U_\theta^\dagger  ) \right)^2,
\end{equation*}
where $U^{j_k} = \partial_{\theta^j_k}U_\theta$.
For convenience we label the Hilbert space on which $U_{\theta^i} \otimes \mathbb{I}_R$ acts as $\mathcal{H}_i$ and so the Hilbert space on which $U_\theta$ acts is $\otimes_{i=1}^n \mathcal{H}_i$. Taking the partial trace over $\mathcal{H}_{B_j} = \otimes_{i \neq j}^n \mathcal{H}_i$, we get 
\begin{eqnarray}
H_{\theta_{mm}}(\rho_0) &=& 4   \tr  \left(( U_{\theta^j}^{j_k}\otimes \mathbb{I} ) \rho^A  (U_{\theta^j}^{j_k \dagger} \otimes \mathbb{I})\right)  \nonumber\\
&+& 4\left(\tr  (U_{\theta^j}^{j_k}  \otimes \mathbb{I} ) \rho^A \left( U_{\theta^j}^\dagger \otimes \mathbb{I} \right) \right)^2 ,
\label{eq:ujuk2}
\end{eqnarray}
where $\rho^A = \tr_{\mathcal{H}_{B_j}} \{ \rho_0 \}$, the reduced state of $\rho_0$. In general, $\rho^A$ is mixed, and so $\rho^A = \sum_{i=1}^{d^2} p_i \rho_i$, where $\rho_i$ are pure and orthogonal. Substituting $\rho^A = \sum_{i=1}^{d^2}  p_i \rho_i$ into (\ref{eq:ujuk2}), we get
\begin{eqnarray*}
H_{\theta_{mm}}(\rho_0) &=& \sum_{i=1}^{d^2}  p_i \bigg( 4   \tr  ( U^{j_k}\otimes \mathbb{I} ) \rho_i  (U^{j_k \dagger} \otimes \mathbb{I} )  \nonumber\\
&+&4  \left(\tr  (U^{j_k}  \otimes \mathbb{I} ) \rho_i ( U^\dagger \otimes \mathbb{I} ) \right)^2\bigg) ,
\label{eq:ujuk2b}
\end{eqnarray*}
and hence by (\ref{eq.sldjk})
\begin{equation*}
H_{\theta_{mm}}(\rho_0) = \sum_{i=1}^{d^2}  p_i ( H_{\theta^j} (\rho_i))_{kk}.
\label{eqn.hjtpihi}
\end{equation*}
Summing over $k$ we get
\begin{eqnarray*}
\sum_k H_{\theta_{m(j,k)m(j,k)}}(\rho_0) &=& \sum_{k=1}^{d^2-1} \sum_{i=1}^{d^2}  p_i  (H_{\theta^j} (\rho_i))_{kk}\\
 &=&  \sum_{i=1}^{d^2}  p_i  \tr \{ H_{\theta^j} (\rho_i)\}.\\
&\leq&  \tr \{ H_{\theta^j} (\rho_{mes})\}.
\label{eqn.hjtpihi2}
\end{eqnarray*}
Summing over $j$ we get (\ref{eqn.htleqhi2}).

\bibliography{PRLbib}

\end{document}